\documentclass{emulateapj}

\shorttitle{Improved Distances to SNe Ia} \shortauthors{Wang et al.}

\def\gsim{\;\lower4pt\hbox{${\buildrel\displaystyle >\over\sim}$}\;}
\def\lsim{\;\lower4pt\hbox{${\buildrel\displaystyle <\over\sim}$}\;}
\def\grls{\;\lower4pt\hbox{${\buildrel\displaystyle >\over <}$}\;}

\begin{document}

\title{Improved Distances to Type Ia Supernovae with
Two Spectroscopic Subclasses}

\author{X. Wang\altaffilmark{1,2}, A. V. Filippenko\altaffilmark{1}, M. Ganeshalingam\altaffilmark{1},
  W. Li\altaffilmark{1}, J. M. Silverman\altaffilmark{1},
  L. Wang\altaffilmark{3}, \\ R. Chornock\altaffilmark{1}, R. J.
  Foley\altaffilmark{1,4,5}, E. L. Gates\altaffilmark{6},
  B. Macomber\altaffilmark{1}, F. J. D. Serduke\altaffilmark{1},
  T. N. Steele\altaffilmark{1}, and D. S. Wong\altaffilmark{1}}

\altaffiltext{1}{Department of Astronomy, University of California,
Berkeley, CA 94720-3411, USA.;wangxf@astro.berkeley.edu.}
\altaffiltext{2}{Physics Department and Tsinghua Center for
Astrophysics (THCA), Tsinghua University, Beijing, 100084, China.}
\altaffiltext{3}{Physics Department, Texas A\&M University, College
Station, TX 77843.} \altaffiltext{4}{Harvard-Smithsonian Center for
Astrophysics, 60 Garden Street, Cambridge, MA, 02138.}
\altaffiltext{5}{Clay Fellow.} \altaffiltext{6}{Lick Observatory,
P.O. Box 85, Mount Hamilton, CA 95140.}

\begin{abstract}
We study the observables of 158 relatively normal Type Ia supernovae
(SNe~Ia) by dividing them into two groups in terms of the expansion
velocity inferred from the absorption minimum of the Si~II
$\lambda$6355 line in their spectra near $B$-band maximum
brightness. One group (``Normal'') consists of normal SNe~Ia
populating a narrow strip in the Si~II velocity distribution, with
an average expansion velocity $\langle v\rangle = 10,600 \pm 400$ km
s$^{-1}$ near $B$ maximum; the other group (``HV'') consists of
objects with higher velocities, $v \gtrsim 11,800$ km s$^{-1}$.
Compared with the Normal group, the HV one shows a narrower
distribution in both the peak luminosity and the luminosity decline
rate $\Delta m_{15}$. In particular, their $B-V$ colors at maximum
brightness are found to be on average redder by $\sim$0.1 mag,
suggesting that they either are associated with dusty environments
or have intrinsically red $B-V$ colors. The HV SNe~Ia are also found
to prefer a lower extinction ratio $R_{V}\approx1.6$ (versus
$\sim$2.4 for the Normal ones). Applying such an
absorption-correction dichotomy to SNe~Ia of these two groups
remarkably reduces the dispersion in their peak luminosity from
0.178 mag to only 0.125 mag.

\end{abstract}

\keywords{cosmology: observations -- distance scale -- dust,
extinction -- supernovae: general}

\section{Introduction}

Type Ia supernovae (SNe~Ia) play important roles in observational
cosmology, with the most compelling evidence for cosmic acceleration
coming from their distance measurements (Riess et~al. 1998;
Perlmutter et~al. 1999). Most SNe~Ia are found to show similar
spectral and photometric behavior (e.g., Suntzeff 1996; Filippenko
1997), suggesting a relatively homogeneous origin --- probably an
accreting carbon-oxygen white dwarf in a binary system (e.g.,
Hillebrandt \& Niemeyer 2000). Yet the presence of some extreme
events such as SN 1991T (Filippenko et~al. 1992a) and SN 1991bg
(Filippenko et~al. 1992b), and the truly peculiar SN 2002cx-like
objects (Li et~al. 2003), may indicate multiple channels producing
the thermonuclear explosion of SNe~Ia.

Diversity is observed even among the relatively normal SNe~Ia. At a
given phase there is a large spread among the absorption blueshifts
of their spectral features (e.g., Benetti et~al. 2005, hereafter
B05), and there are also differences in line strengths. The scatter
comes primarily from objects showing high photospheric velocities.
Representative examples are SNe 2002bo, 2002dj, 2004dt, and 2006X
(Benetti et~al. 2004; Pignata et~al. 2008; L. Wang et al. 2006; Wang
et~al. 2008a, hereafter W08a); the Si~II velocities near $B$-band
maximum light are higher than those of the normal SNe~Ia by
$\sim$2500--5500 km s$^{-1}$. Such SNe~Ia are defined as a subclass
in terms of the temporal velocity gradient of Si~II (B05), and also
grouped by Branch et~al. (2006) according to the equivalent width
(EW) of the Si~II $\lambda$6355 and Si~II $\lambda$5972 absorptions
lines. The origin of the high expansion velocities is debated, with
a conventional mechanism being a density/abundance enhancement in
the outer ejecta (e.g., Tanaka et al. 2008). A statistical study of
the observables may allow one to penetrate deeper into their
explosion physics and understand their impact on the use of SNe~Ia
to measure the properties of dark energy.

The main thrust of this {\it Letter} is to show that SNe~Ia with
high expansion velocities may have a lower extinction ratio with
respect to the normal ones. Using two different values of $R_{V}$ in
their brightness corrections, the utility of SNe Ia for cosmological
distance determinations can be substantially increased.

\section{Spectroscopic Classification of SNe~Ia}

The sample in our study includes most of the SNe~Ia available in the
literature and in our
database\footnote{Peculiar objects such as SN 1991T and SN 1991bg
were not included in this sample. The criteria used to identify the
91T-like objects include weak Si~II absorption and prominent
Fe~III lines in the near-maximum spectra, while identification of
the 91bg-like events relies on the presence of obvious Ti~II
absorption or a very deep Si~II $\lambda$5972 feature (the Si
line-depth ratio [$\Re$(Si~II); Nugent et al. 1995] is larger than
0.50). The list of the peculiar events can be provided upon request.};
it consists of 158 relatively normal SNe Ia
with good photometry and generally at least one spectrum within one
week after $B$ maximum. The spectral data are primarily from
Matheson et~al. (2008, hereafter M08) and our own database
(Silverman et~al., in prep., hereafter S09). For a few objects, the
spectral parameters were also taken from B05, Branch et al. (2009),
and the IAU Circulars. Table 1 lists the observed parameters as well
as the classifications of the SNe~Ia.

\subsection{Expansion Velocity from Si~II$\lambda$6355}

Si~II $\lambda$6355 is one of the strongest features in
optical/near-infrared spectra of SNe~Ia; the blueshift of its
absorption minimum has often been used to diagnose the diversity
among SNe~Ia. B05 distinguished SNe~Ia having a high Si~II temporal
velocity gradient (HVG) from those with a low Si~II temporal velocity gradient
(LVG). However, relatively few objects have the multi-epoch spectra
(covering phases from maximum brightness to a few weeks thereafter)
necessary for measurement of such a velocity gradient. As the HVG
SNe~Ia generally have faster expansion velocities than the LVG ones,
we may nominally divide the SN~Ia sample into ``Normal'' and ``HV''
groups according to the observed velocity of Si~II $\lambda$6355.

\begin{figure}
\figurenum{1}
\vspace{-0.5cm} \hspace{-0.5cm}
\includegraphics[angle=0,width=97mm,height=78mm]{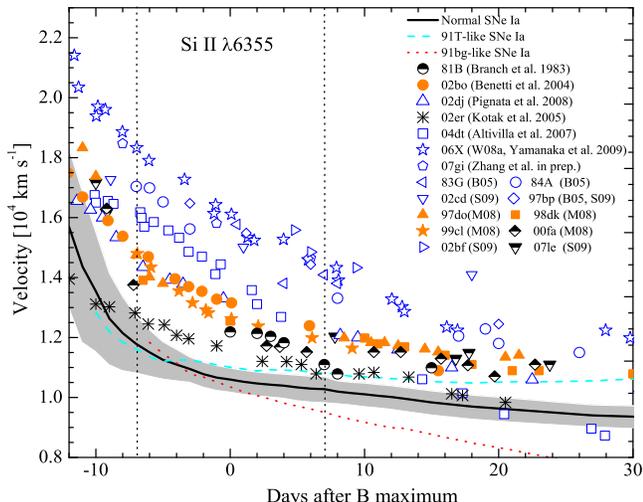}
\vspace{-1.0cm} \caption{Temporal evolution of the expansion
velocity inferred from the blueshift of the Si~II $\lambda$6355
absorption minimum. The solid line shows the mean evolution obtained
from 10 well-observed Normal SNe~Ia, and the gray region represents
the 1$\sigma$ uncertainty; the dashed and dotted lines illustrate
the evolution of the mean velocity for SN 1991T-like and SN
1991bg-like events, respectively (the data sources are M08 and S09).
Overplotted is part of the HV sample, while SN 2002er may be a
transitional object linking the HV and Normal groups.} \label{fig-1}
\end{figure}

Based on the Si~II velocity distribution of 10 well-observed Normal
SNe~Ia\footnote{The sample includes SNe 1989B, 1994D, 1997dt,
1998aq, 1998bu, 1999ee, 2003cg, 2003du, 2004eo, and 2005cf (Barbon
et~al. 1990; Patat et~al. 1996; M08; Branch et~al. 2003; Hamuy
et~al. 2002; Elisa-Rosa et~al. 2006; Stanishev et~al. 2007;
Pastorello et~al. 2007; Garavini et~al. 2007; Wang et~al. 2009).}, we
derive their mean velocity from $t = -12$~day to +30~day. The
evolution of the mean velocity is shown in Figure 1, where the gray
area indicates 1$\sigma$ uncertainty obtained through Monte Carlo
simulations. After maximum brightness, the velocity evolves nearly
in a linear fashion, with a gradient of about 40 km s$^{-1}$
day$^{-1}$ and a typical scatter $\sim \pm$400 km s$^{-1}$. The
large scatter shown before $t \approx -7$~day is caused by detached
HV features at the earliest epochs. In comparison with the lower
velocity and homogeneous distribution seen in the Normal SNe~Ia, the
expansion velocities of the HV SNe~Ia are higher but more scattered,
with a faster decay. The highest contrast in the Si~II velocity
between these two groups occurs within one week from the maximum
brightness. After $t \approx +7$~day, the velocities of some HV SNe~Ia
are comparable to those of the Normal ones. We thus use the velocity
measured within one week from the maximum to subclassify our sample;
the value obtained with the spectrum closer to $B$ maximum is
adopted when multi-epoch measurements are available. By applying a
3$\sigma$ selection criterion, 55 of the 158 objects were identified
as HV SNe~Ia. We note that there is no sharp division between these
two groups when the velocity approaches a lower value (see also Fig.
2), so blending could occur to some extent.

\subsection{Equivalent Width of Si~II $\lambda$6355}

An alternative way to quantify the diversity of SNe~Ia is through
the line-strength ratio (e.g., Nugent et al. 1995) or the equivalent
width (EW) of some features (e.g., Hachinger et~al. 2006). Based on
the EW of the absorption near Si II$\lambda$5972 and Si
II$\lambda$6355, Branch et~al. (2006, 2009) suggest dividing the
SN~Ia sample into four groups: cool (CL), shallow silicon (SS), core
normal (CN), and broad line (BL); their CL and SS groups consist
mainly of peculiar events such as SN 1991bg and SN 1991T,
respectively. Compared with the Branch CN SNe~Ia, our definition of
the Normal sample is wider, while their BL objects overlap well with
our HV sample.

\begin{figure}
\figurenum{2} \vspace{-0.65cm}\hspace{-0.7cm}
\includegraphics[angle=0,width=95mm,height=80mm]{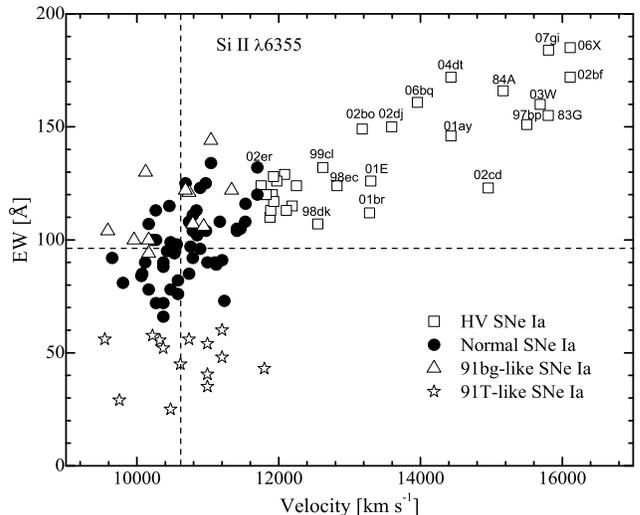}
\vspace{-1.0cm} \caption{EW of Si~II $\lambda$6355 absorption versus
the expansion velocity, measured from spectra taken within $\pm$3
day of $B$ maximum, for SNe~Ia of different subtypes. The vertical
and horizontal lines mark the respective locations of the mean
velocity and EW of the Normal SNe~Ia.} \label{fig-2}
\end{figure}

Figure 2 shows a plot of the EW versus the velocity of Si~II
$\lambda$6355 absorption, obtained for SNe~Ia with spectra near $B$
maximum. One can see that the Si~II absorption is generally strong
in the HV subclass, typically with $EW \gtrsim 100$~\AA, and the
strength of the absorption correlates with the expansion velocity.
In principle, strong absorption at high velocity can be caused by an
enhancement of abundance or density in the outermost layers, perhaps
due to an extended burning front (B05) or an interaction with
circumstellar material (e.g., Gerardy et~al. 2004). The continuous
distribution of the EW and velocity between the HV and Normal SNe~Ia
suggests that such an enhancement process occurs to different
degrees with considerable probability. This could be interpreted as
a line-of-sight effect if the high velocities are caused by
aspherical structures, such as a thick torus, as evidenced by the
intrinsically large polarization detected for HV objects (e.g., L.
Wang et~al. 2006).

\section{Photometric Properties of \\ the HV and Normal SNe~Ia}

Given the spectroscopic diversity of the HV and the Normal groups
addressed in \S2, it is interesting to compare their photometric
behaviors. The peak magnitude and decline rate $\Delta m_{15}$
(Phillips 1993) are taken from the literature (e.g., Reindl et~al.
2005; Hicken et~al. 2008) or estimated from our own photometry
(Ganeshalingam et~al., in prep.) obtained primarily with the Lick
Observatory 0.76-m Katzman Automatic Imaging Telescope (KAIT;
Filippenko et al. 2001) and the 1-m Nickel reflector. Distances to
the SNe are computed by using the redshifts of their host galaxies,
in the reference frame of the 3~K cosmic microwave background
radiation for samples at redshift $z \gtrsim 0.01$ or corrected to a
self-consistent Virgocentric infall of 220 km s$^{-1}$ for closer
ones, with a Hubble constant $H_{0} = 70.5$ km s$^{-1}$ Mpc$^{-1}$
(Komatsu et~al. 2009). A peculiar-velocity component of 300 km
s$^{-1}$ is included in the distance-modulus uncertainties. Cepheid
distances are adopted whenever available (see X. Wang et~al. 2006,
and references therein).

\subsection{Luminosity and the Secondary Parameters}

Shown in Figure 3 are the distributions of the $V$-band peak
absolute magnitudes $M^{V}_{\rm max}$, the $B_{\rm max} - V_{\rm
max}$ color, $\Delta m_{15}$, and the morphological T-type of the
host galaxy (de Vaucouleurs et al. 1991, hereafter RC3) for the HV
and Normal SNe~Ia. Both $M^{V}_{\rm max}$ and $B_{\rm max} - V_{\rm
max}$ were $K$-corrected (e.g., Jha et al. 2007) and dereddened by
the Galactic reddening using the full-sky maps of dust infrared
emission (Schlegel et al. 1998).

\begin{figure}
\figurenum{3} \vspace{-0.5cm}\hspace{-0.8cm}
\includegraphics[angle=0,width=100mm,height=80mm]{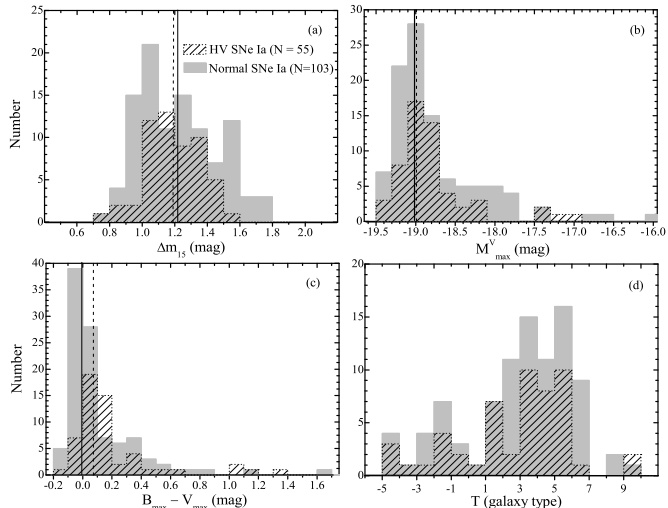}
\vspace{-1.0cm} \caption{Histogram distribution of $\Delta m_{15}$,
the $V$-band absolute magnitudes, the $B_{\rm max} - V_{\rm max}$
colors, and the T-types of the host galaxies. Two vertical lines in
panels (a), (b), and (c) mark the respective mean values of the
observables for the Normal (solid) and HV (dashed) SNe~Ia with
$B_{\rm max} - V_{\rm max} < 0.20$ mag.} \label{fig-3}
\end{figure}

As can be seen from Figure 3a, the decline rate of the HV SNe~Ia
exhibit a narrower distribution relative to the Normal group.
Although they include events with small $\Delta m_{15}$, no HV
objects were found at $\Delta m_{15} > 1.6$. A similar narrow
distribution is seen in their $V$-band absolute magnitudes (Fig.
3b), with exceptions for a few heavily reddened objects on the faint
side. Restricting the sample to those with $B_{\rm max} - V_{\rm
max} < 0.20$ mag, the mean value of $M^{V}_{\rm max}$ as well as that of
$\Delta m_{15}$ is found to be comparable for the HV and Normal
SNe~Ia. Despite these similarities, the $B_{\rm max} - V_{\rm max}$
colors of the two groups show noticeable differences (see Fig. 3c),
with the average value of the HV group being redder by $\sim$0.08
mag. This difference increases to 0.10 mag by further restricting
the subsamples to those with $1.0 \lesssim \Delta m_{15} \lesssim
1.5$. In addition, the frequency of the Normal SNe~Ia at the bluer
end ($B_{max} - V_{max}<$0) is obviously higher than that of the HV
objects, e.g., 43.6\% vs. 14.5\%. This indicates that the HV SNe~Ia
may preferentially occur in dusty environments, or they have
intrinsically red colors (see discussion in \S 4). On the other
hand, the morphological distributions of the host galaxies for these
two groups do not show significant differences, perhaps suggesting
that the color differences might not be caused by dust on large
scales.

Although our analysis involves a large sample, we caution that the
distributions of the above observables may suffer from an
observational bias inherited in the observed sample. Further studies
will be needed to determine whether this is indeed the case.

\subsection{Dust Absorption and Luminosity Standardization}

Dust absorption may be one of the main uncertainties in the
brightness corrections of SNe~Ia, depending not only on knowledge of
the reddening $E(B-V)$ but also on the properties of the dust.
Infrared photometry, together with optical data, would set better
constraints on $R_{V}$ (defined to be $A_V/E(B-V)$) on an
object-by-object basis, but it was not available for most of our
sample. Given the sample size, we attempt to quantify the average
extinction ratio $R_{V}$ for SNe~Ia in a statistical way.

With the known empirical relations between the intrinsic colors and the
decline rate $\Delta m_{15}$ (Wang et al. 2009), we derive the
reddening for our sample from the $B_{\rm max} - V_{\rm max}$ color
and that measured at 12 days past the $B$ maximum (X. Wang et~al.
2005). To maintain consistency in our reddening determination, we
did not use the tail color (Phillips et~al. 1999) or the color at $t
= 35$~day after $B$ maximum (Jha et~al. 2007) owing to the possible
abnormal color evolution of HV SNe~Ia in the nebular phase (W08a).
The host-galaxy reddening $E(B-V)_{\rm host}$ was taken to be the
weighted average of the determinations by the above two methods, and
the negative values were kept as measured. We note, however, that
part of the reddening derived for the HV objects would be biased if
their intrinsic color-$\Delta m_{15}$ relations were different from
those defined by the normal ones. Thus, the inferred value of $R_{V}$
for them (discussed below) may not represent the true extinction
ratio.

In Figure 4, the absolute peak magnitudes $M^{V}_{\rm max}$,
corrected for the dependence on $\Delta m_{15}$ using a relation
derived from the low-reddening subsample, is plotted versus
$E(B-V)_{\rm host}$. One can see that the HV group follows a
relation significantly different from that for the Normal group.
Assuming that the correlation is governed by dust absorption, the
effective $R_{V}$ values for these two groups are $1.57 \pm 0.08$
(HV) and $2.36 \pm 0.09$ (Normal). The objects with $z < 0.010$ but
without Cepheid distances were not included in the fit\footnote{SN
2006lf was excluded from the fit. After correction for the Galactic
reddening $E(B - V)_{\rm Gal}$ = 0.97 mag with $R_{V}$ = 3.1, the
$V$-band luminosity is found to be higher than that of the typical
SN Ia by about 0.7 mag, possibly indicating a large uncertainty in
the correction for Galactic reddening.}. We note that SN 1996ai and
SN 2003cg may have $R_{V}$ close to 1.9, perhaps due to abnormal
interstellar dust, although their distances have large
uncertainties. It is not clear whether they are outliers, or a low
$R_{V}$ value is common for extremely reddened SNe~Ia. Nevertheless,
the slopes of the fits to the HV and Normal SNe~Ia are still clearly
different even if all the events with $E(B-V)_{\rm host} > 0.50$ mag
are excluded. To examine the robustness of our analysis, we separate
our sample using a 4$\sigma$ selection criterion. This gives
best-fit values of $R_{V}$ as $1.53 \pm 0.07$ for the 25 hubble-flow
HV SNe~Ia and $2.32 \pm 0.08$ for the remaining 100 objects,
consistent with the values determined from the 3$\sigma$ subsamples.

\begin{figure}
\figurenum{4} \vspace{-0.3cm}\hspace{-0.7cm}
\includegraphics[angle=0,width=100mm,height=80mm]{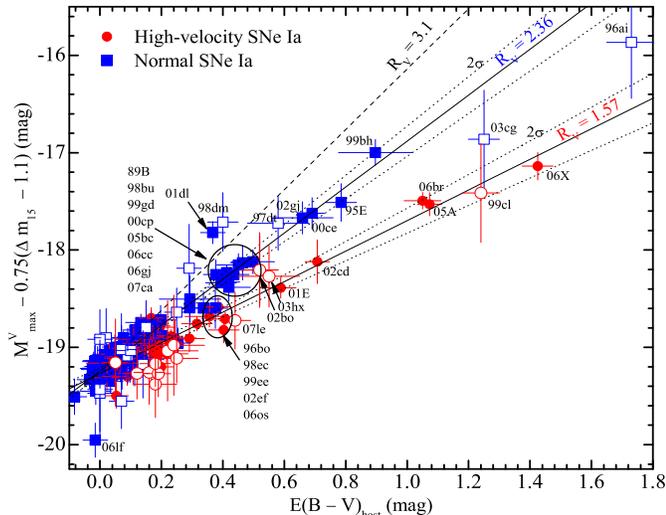} \vspace{-1.0cm}
\caption{The $\Delta m_{15}$-corrected absolute $V$ mag at maximum
brightness versus the host-galaxy reddening. The filled symbols are
SNe with $z \gtrsim 0.01$ or Cepheid-based distances, and the open
symbols are nearby objects that were not included in the fit. The
two solid lines show the best-fit $R_{V}$ for SNe in the HV and
Normal groups, with dotted lines indicating 2$\sigma$ uncertainties.
The dashed line represents the Milky Way reddening law.}
\label{fig-4}
\end{figure}

The corresponding regression of $M^{V}_{\rm max}$ with two
variables, $\Delta m_{15}$ and $E(B-V)_{\rm host}$ takes the form
\begin{equation} M^{V}_{max} = M_{\rm zp}+ \alpha(\Delta m_{15} -
1.1) + R_{V}E(B-V)_{\rm host},
\end{equation}
where $M_{\rm zp}$ represents the mean absolute magnitude, corrected
for the reddening in the Milky Way and the host galaxy, and
normalized to $\Delta m_{15}$ = 1.1. For the HV and Normal SNe Ia,
one obtains

\begin{equation}
\begin{array}{rl}
{\rm Normal}: \alpha = 0.77\pm0.06, R_{V} = 2.36\pm0.07,\\
M_{\rm zp} = -19.26\pm0.02, N = 83, \sigma = 0.123;
\end{array}
\end{equation}

\begin{equation}
\begin{array}{rl}
{\rm HV:} \alpha = 0.75\pm0.11, R_{V} = 1.58\pm0.07, \\
M_{\rm zp} = -19.28\pm0.03, N = 42, \sigma = 0.128.
\end{array}
\end{equation}

The improved solutions are quite close to the provisional values.
The resulting $M^{V}_{\rm max}$$- \Delta m_{15}$ relation does not
show a significant difference between the HV and Normal groups, and
both slopes are found to be in accord with that adopted in the
earlier analysis.

After accounting for the dependence on the two variables $\Delta
m_{15}$ and $E(B-V)_{\rm host}$, we find the luminosity scatter to
be 0.128 mag for the HV group and 0.123 mag for the Normal one. A
two-component fit to these two groups having different values of
$R_{V}$ yields a luminosity scatter 0.125 mag. Assuming a single
best-fit $R_{V}=1.85$, however, the luminosity scatter increases to
0.178 mag. We further performed the analysis using a subsample with
$1.0 \lesssim \Delta m_{15} \lesssim 1.5$, which yields the best-fit
$R_{V}$ as 2.28$\pm$0.09 (Normal) and 1.54$\pm$0.08 (HV),
respectively. This demonstrates that our results are not affected by
objects at the extreme ends. Replacing the reddening term in Eq.(1)
with the $B_{\rm max} - V_{\rm max}$ color, the best-fit slopes are
found to be 2.29$\pm$0.08 (Normal) and 1.55$\pm$0.06 (HV). Such a
dichotomy is also required for the pure color-term corrections,
which can decrease the luminosity scatter from 0.177 mag to 0.136
mag. We thus propose that applying two different $R_{V}$ values to
the brightness corrections of SNe~Ia is potentially beneficial,
significantly decreasing the uncertainties in their distance
measurements.

\section{Discussion and Conclusions}

We demonstrate that the standardization of SNe Ia can be noticeably
improved by separating them into Normal and HV groups using a
spectroscopic criterion (the blueshift of the absorption minimum of
Si II $\lambda$6355). The main advantage of such a distinction is that
SNe Ia of these two groups may have different extinction laws,
though the $R_{V}$ inferred for the HV SNe Ia might be partially
biased due to the possibly different colors. By contrast with the
Normal SNe~Ia, the HV objects are found to have red $B - V$ color.
This difference could be due to reddening, intrinsic color variance,
observational bias, or a combination of these factors.

In the dust scenario, an additional absorption component, such as
circumstellar (CS) dust (L. Wang et al. 2005), may be required to
account for the paucity of HV SNe~Ia at the bluest end (see Table 1
and Fig. 4). Tantalizing evidence for the presence of CS material
around HV SNe~Ia is provided by the detection of variable Na~I~D
absorption lines in SN 2006X and 1999cl (Patat et~al. 2007; Wang
et~al. 2008b; Blondin et~al. 2009). The lower value of $R_{V}$ might
be naturally explained by multiple scattering in the CS dust shell
(Goobar 2008).

As an alternative to reddening, the red $B - V$ color seen in the HV
SNe~Ia may be intrinsic, at least partially. Possible causes include
the metallicity (Dominguez et al. 2001; Timmes et al. 2003) and a
different extent of the burning front in the outer layers (Benetti
et al. 2004). Increasing the metallicity of the progenitor could
lead to a slightly redder $B-V$ color due to line blanketing; on the
other hand, the increased opacity, as a result of the density
enhancement in the outer ejecta, may also lead to a red color
because of a low photospheric temperature at the earlier phases. A
quantitative analysis would help determine whether the differences
in the color as well as the inferred $R_{V}$ values could be
reproduced by the above two scenarios.

We emphasize that regardless of the origin of this color difference,
the improvement in the distances to SNe~Ia by applying two values of
$R_{V}$ will persist, e.g., with an uncertainty decreasing from
$\sim$9\% to 6\%. Analysis of the impact of our findings on current
cosmological studies will be presented in a forthcoming paper.

\acknowledgements

We thank Mark Phillips for allowing us to use the photometric
parameters of SN 2005A before publication, D. C. Leonard for useful
discussions, and the Lick Observatory staff for their assistance
with the observations. C. V. Griffith, J. J. Kong, N. Lee, and E.
Miller helped maintain and improve the SN Database of A.V.F.'s
supernova group at UC Berkeley. We are grateful to many students,
postdocs, and other collaborators who have contributed to
observations and reductions of our SN spectra and images over the
past two decades, especially C. Anderson, A. J. Barth, L.-B.
Desroches, G. Foster, B. Grigsby, N. Joubert, D. C. Leonard, T.
Matheson, M. Moore, M. Papenkova, S. Park, B. Swift, T. Pritchard,
and D. Winslow. This group is supported by NSF grant AST--0607485,
the TABASGO Foundation, US Department of Energy SciDAC grant
DE-FC02-06ER41453, and US Department of Energy grant
DE-FG02-08ER41563. We are also grateful to the National Natural
Science Foundation of China (10673007), and the China-973 Program
2009CB824800. The work of L. Wang is supported by NSF grant
AST-0708873. KAIT and its ongoing operation were made possible by
donations from Sun Microsystems, Inc., the HP Company, AutoScope
Corporation, Lick Observatory, the NSF, the University of
California, the Sylvia \& Jim Katzman Foundation, and the TABASGO
Foundation. We made use of the NASA/IPAC Extragalactic Database
(NED), which is operated by the Jet Propulsion Laboratory,
California Institute of Technology, under contract with NASA.

\clearpage


\clearpage

\LongTables
\begin{deluxetable}{lccccccc}
\tablewidth{0pt} \tabletypesize{\scriptsize} \tablecaption{Observed
Parameters of the SN Ia Sample$^{\ast}$.}
\tablehead{
\colhead{SN} &
 \colhead{$v_{3K/220}$(km s$^{-1}$)} &
 \colhead{$M^{V}_{\rm max}$(mag)}&
 \colhead{$\Delta m_{15}$$^{\ddagger}$(mag)} &
 \colhead{$B_{\rm max} - V_{\rm max}$(mag)} &
 \colhead{$E(B-V)_{\rm host}$(mag)} & \colhead{T(RC3)} &
 \colhead{SN type}
}
\startdata

1989B$\dagger$  & 549  & -18.15(0.15) & 1.35(0.05) & 0.32(0.07) & 0.39(0.05) & 3 & N \\
1990N$\dagger$  & 1179 & -19.09(0.14) & 1.05(0.05) & 0.00(0.05) & 0.11(0.04) & 4 & N \\
1994ae$\dagger$ & 1575 & -19.35(0.15) & 0.89(0.05) &-0.05(0.05) & 0.04(0.04) & 5 & N \\
1994D  & 1179 & -19.26(0.57) & 1.31(0.05) &-0.08(0.05) &-0.04(0.04) & -2 & N \\
1995D  & 2129 & -19.18(0.31) & 1.02(0.05) &-0.02(0.05) & 0.09(0.04) & -1 & N \\
1995E  & 3496 & -17.48(0.19) & 1.17(0.07) & 0.70(0.05) & 0.78(0.05) & 3 & N \\
1996ai & 1174 & -15.94(0.57) & 1.00(0.08) & 1.69(0.05) & 1.75(0.08) & 4 & N \\
1996X  & 2120 & -19.36(0.31) & 1.34(0.05) &-0.01(0.05) & 0.06(0.04) & -4 & N \\
1997dt & 2356 & -17.72(0.29) & 1.11(0.10) & 0.50(0.10) & 0.59(0.09) & 5 & N \\
1997E  & 4055 & -18.73(0.18) & 1.42(0.05) & 0.02(0.11) & 0.11(0.09) & -2 & N \\
1998aq$\dagger$ & 1514 & -19.25(0.14) & 1.05(0.05) &-0.11(0.04) &-0.02(0.04) & 2 & N \\
1998bu$\dagger$ & 810  & -18.36(0.14) & 1.06(0.04) & 0.33(0.04) & 0.42(0.04) & 2 & N \\
1998dm & 1976 & -17.86(0.33) & 0.91(0.05) & 0.30(0.04) & 0.39(0.05) & 5 & N \\
1998eg & 7056 & -18.84(0.11) & 1.17(0.10) & 0.00(0.08) & 0.12(0.06) & 6 & N \\
1999bh & 5295 & -16.64(0.13) & 1.57(0.20) & 0.87(0.09) & 0.90(0.12) & 3 & N \\
1999cp & 3112 & -19.28(0.21) & 1.02(0.03) & 0.01(0.03) & 0.07(0.04) & 6 & N \\
1999dg & 6782 & -18.97(0.10) & 1.45(0.08) &-0.05(0.06) & 0.02(0.04) & -2& N \\
1999ee & 3163 & -18.70(0.21) & 0.96(0.05) & 0.29(0.04) & 0.38(0.04) & 4 & N \\
1999ej & 4630 & -18.70(0.14) & 1.54(0.05) & 0.00(0.04) & 0.03(0.04) & 0 & N \\
1999gd & 5775 & -18.10(0.12) & 1.13(0.10) & 0.41(0.08) & 0.50(0.07) & 2 & N \\
2000ca & 7080 & -19.39(0.10) & 0.96(0.05) &-0.09(0.04) & 0.02(0.04) & 5 & N \\
2000ce & 4940 & -17.74(0.17) & 1.00(0.10) & 0.59(0.16) & 0.66(0.12) & 3 & N \\
2000cf & 10930& -18.90(0.07) & 1.43(0.05) &-0.02(0.04) & 0.04(0.05) & \nodata & N \\
2000cn & 6958 & -18.53(0.10) & 1.57(0.05) & 0.18(0.04) & 0.20(0.05) & 6 & N \\
2000cp & 10341& -18.16(0.12) & 1.10(0.10) & 0.37(0.14) & 0.45(0.07) & 2 & N \\
2000cw & 8692 & -18.77(0.08) & 1.26(0.05) & 0.01(0.03) & 0.14(0.08) & 4 & N \\
2000dk & 4932 & -18.93(0.14) & 1.66(0.03) &-0.01(0.04) &-0.01(0.04) & -5 & N \\
2000dm & 4393 & -18.88(0.15) & 1.45(0.06) &-0.05(0.04) &-0.01(0.04) & 2 & N \\
2000dn & 9248 & -18.99(0.08) & 1.04(0.06) & 0.00(0.04) & 0.11(0.04) & -1 & N \\
2000dr & 5332 & -18.52(0.13) & 1.73(0.05) & 0.10(0.04) & 0.05(0.04) & -1 & N \\
2001ba & 9194 & -19.31(0.08) & 0.98(0.05) &-0.10(0.04) & 0.01(0.04) & 3.7 & N \\
2001bf & 4556 & -19.38(0.14) & 0.79(0.04) & 0.00(0.04) & 0.10(0.04) & -3 & N \\
2001bg & 2273 & -18.64(0.29) & 1.10(0.04) & 0.15(0.04) & 0.26(0.04) & 3 & N \\
2001bt & 4332 & -18.84(0.15) & 1.25(0.05) & 0.15(0.07) & 0.25(0.04) & 4 & N \\
2001cj & 7507 & -19.20(0.09) & 0.93(0.04) &-0.08(0.04) & 0.01(0.04) & 6 & N \\
2001ck & 10583& -19.10(0.07) & 0.99(0.04) &-0.04(0.04) & 0.06(0.04) & 3 & N \\
2001cp & 6714 & -19.29(0.10) & 0.85(0.05) & 0.01(0.05) & 0.13(0.04) & 4 & N \\
2001dl & 5875 & -18.01(0.11) & 0.85(0.05) & 0.26(0.04) & 0.37(0.04) & 8 & N \\
2001el & 1030 & -18.12(0.65) & 1.16(0.05) & 0.06(0.04) & 0.20(0.06) & 6 & N \\
2001ep & 3881 & -18.88(0.17) & 1.43(0.05) & 0.02(0.03) & 0.14(0.06) & 3 & N \\
2001fe & 4344 & -19.30(0.15) & 1.03(0.10) & 0.03(0.04) & 0.11(0.05) & 1 & N \\
2001fh & 3630 & -19.25(0.18) & 1.45(0.04) &-0.16(0.05) &-0.08(0.04) & 3 & N \\
2002aw & 7850 & -19.04(0.09) & 1.22(0.08) & 0.05(0.04) & 0.14(0.07) & 3 & N \\
2002cr & 3112 & -19.07(0.21) & 1.21(0.05) &-0.01(0.03) & 0.10(0.05) & 6 & N \\
2002dp & 3133 & -18.82(0.21) & 1.25(0.05) & 0.05(0.04) & 0.16(0.04) & 5 & N \\
2002eb & 7910 & -19.19(0.09) & 0.92(0.03) &-0.07(0.04) & 0.07(0.04) & 4 & N \\
2002el & 8337 & -19.16(0.08) & 1.25(0.08) &-0.03(0.06) & 0.07(0.04) &-3 & N \\
2002er & 2804 & -18.92(0.23) & 1.32(0.03) & 0.12(0.03) & 0.23(0.04) & 1 & N \\
2002fk & 2007 & -18.95(0.33) & 1.06(0.03) &-0.1 (0.04) & 0.00(0.04) & 4 & N \\
2002ha & 3913 & -18.96(0.17) & 1.34(0.07) &-0.08(0.03) &-0.02(0.04) & 4 & N \\
2002he & 7447 & -18.92(0.09) & 1.32(0.06) & 0.04(0.04) & 0.10(0.04) &-5 & N \\
2002hu & 11460& -19.36(0.06) & 1.05(0.07) &-0.11(0.05) & 0.02(0.05) & 5 & N \\
2002hw & 4896 & -18.03(0.14) & 1.57(0.10) & 0.39(0.08) & 0.42(0.07) & 5 & N \\
2002jg & 4158 & -17.31(0.16) & 1.52(0.05) & 0.64(0.04) & 0.69(0.04) & 5 & N \\
2003cg & 1340 & -16.77(0.50) & 1.22(0.10) & 1.17(0.05) & 1.23(0.05) & 4 & N \\
2003du & 2307 & -19.01(0.28) & 0.97(0.06) &-0.09(0.04) & 0.02(0.04) & 8 & N \\
2003gt & 4407 & -19.08(0.15) & 1.01(0.03) & 0.01(0.04) & 0.10(0.04) & 2 & N \\
2003he & 7287 & -18.92(0.09) & 1.01(0.05) & 0.05(0.04) & 0.15(0.04) & 4 & N \\
2003hv & 1525 & -19.15(0.43) & 1.39(0.07) &-0.10(0.06) & 0.00(0.04) &-2 & N \\
2003kf & 2085 & -19.09(0.32) & 1.01(0.08) & 0.02(0.09) & 0.06(0.05) & 3 & N \\
2004at & 7082 & -19.23(0.09) & 1.10(0.03) &-0.10(0.04) &-0.02(0.04) & 3 & N \\
2004bd & 2965 & -17.99(0.24) & 1.78(0.10) & 0.37(0.14) & 0.29(0.10) & 1 & N \\
2004bw & 6553 & -19.03(0.10) & 1.27(0.06) &-0.03(0.04) & 0.02(0.04) & 5 & N \\
2004eo & 4421 & -18.95(0.15) & 1.43(0.03) &-0.01(0.03) & 0.06(0.04) & 2 & N \\
2004ey & 4438 & -19.17(0.15) & 1.17(0.07) &-0.09(0.06) & 0.00(0.04) & 5 & N \\
2004fz & 4935 & -19.28(0.14) & 1.32(0.07) &-0.07(0.06) &-0.01(0.05) & 5 & N \\
2004gs & 8249 & -18.32(0.08) & 1.67(0.10) & 0.11(0.04) & 0.13(0.05) &-2? & N \\
2004S  & 2955 & -18.99(0.22) & 1.10(0.07) &-0.04(0.06) & 0.07(0.04) & 5 & N \\
2005am & 2308 & -18.97(0.29) & 1.52(0.08) & 0.09(0.06) & 0.12(0.04) & 1 & N \\
2005bc & 3831 & -17.82(0.18) & 1.52(0.05) & 0.43(0.07) & 0.46(0.04) & 3 & N \\
2005bo & 4504 & -18.58(0.15) & 1.12(0.07) & 0.20(0.06) & 0.29(0.05) & 2 & N \\
2005cf & 2112 & -19.13(0.31) & 1.05(0.03) &-0.02(0.05) & 0.11(0.03) &-2 & N \\
2005de & 4460 & -18.70(0.15) & 1.19(0.05) & 0.04(0.04) & 0.17(0.05) &\nodata & N \\
2005el & 4465 & -19.09(0.15) & 1.24(0.03) &-0.08(0.03) &-0.02(0.03) &-2 & N \\
2005hc & 13500& -19.04(0.06) & 0.98(0.05) &-0.06(0.06) & 0.08(0.04) &\nodata & N \\
2005iq & 9879 & -18.91(0.07) & 1.09(0.08) &-0.03(0.04) & 0.08(0.04) & 1 & N \\
2005kc & 4167 & -18.49(0.17) & 1.24(0.10) & 0.25(0.07) & 0.34(0.05) & 2 & N \\
2005ki & 6111 & -19.14(0.11) & 1.22(0.05) &-0.03(0.05) & 0.03(0.04) &-3 & N \\
2005lz & 13200& -18.90(0.07) & 1.37(0.10) & 0.07(0.05) & 0.12(0.05) &\nodata & N \\
2005ms & 7771 & -19.07(0.09) & 0.83(0.07) &-0.02(0.05) & 0.13(0.04) &\nodata & N \\
2005na & 8044 & -19.24(0.09) & 1.23(0.07) &-0.14(0.05) &-0.03(0.04) & 1 & N \\
2006ax & 5387 & -19.36(0.12) & 1.08(0.04) &-0.05(0.03) & 0.03(0.04) & 3.5 & N \\
2006az & 9435 & -19.11(0.08) & 1.30(0.05) &-0.07(0.04) & 0.00(0.04) &\nodata & N \\
2006cc & 9793 & -18.26(0.07) & 1.06(0.05) & 0.35(0.03) & 0.41(0.04) &\nodata & N \\
2006D  & 2680 & -18.82(0.25) & 1.38(0.09) & 0.03(0.06) & 0.11(0.05) & 2 & N \\
2006dm & 6240 & -18.77(0.11) & 1.56(0.05) & 0.01(0.04) & 0.04(0.03) & 5 & N \\
2006en & 9234 & -18.90(0.08) & 0.99(0.06) & 0.01(0.05) & 0.11(0.04) & 6 & N \\
2006gj & 8310 & -18.03(0.09) & 1.40(0.17) & 0.36(0.09) & 0.44(0.05) & 2 & N \\
2006hb & 4601 & -18.64(0.15) & 1.67(0.08) & 0.06(0.06) & 0.08(0.04) &-3 & N \\
2006kf & 6240 & -18.86(0.11) & 1.52(0.07) &-0.03(0.06) & 0.02(0.04) &\nodata & N \\
2006lf & 3889 & -19.92(0.18) & 1.15(0.05) &-0.04(0.06) &-0.02(0.04) &\nodata & N \\
2006N  & 4278 & -18.83(0.16) & 1.55(0.07) &-0.01(0.07) &-0.02(0.04) &\nodata & N \\
2006ob & 17477& -18.84(0.05) & 1.60(0.12) & 0.05(0.04) & 0.05(0.05) & 3 & N \\
2006S  & 9875 & -18.98(0.07) & 0.93(0.04) & 0.05(0.03) & 0.17(0.04) & \nodata & N \\
2006td & 4512 & -18.41(0.15) & 1.50(0.12) & 0.12(0.05) & 0.20(0.05) & \nodata & N \\
2007af & 1918 & -19.08(0.34) & 1.22(0.05) & 0.04(0.03) & 0.15(0.04) & 6 & N \\
2007bc & 6555 & -18.95(0.11) & 1.30(0.08) &-0.04(0.06) & 0.05(0.04) & 1 & N \\
2007bm & 1996 & -18.22(0.33) & 1.20(0.09) & 0.46(0.06) & 0.55(0.04) & 5 & N \\
2007ca & 4521 & -18.39(0.15) & 0.91(0.05) & 0.29(0.04) & 0.38(0.04) & 5 & N \\
2007ci & 5762 & -18.71(0.12) & 1.77(0.09) & 0.02(0.06) &-0.02(0.04) &-5 & N \\
2007F  & 7244 & -19.15(0.09) & 0.95(0.07) &-0.05(0.04) & 0.06(0.04) & 6 & N \\
2007O  & 10969& -19.17(0.08) & 1.12(0.08) &-0.04(0.07) & 0.05(0.04) & 5 & N \\
2008bf & 6647 & -19.18(0.11) & 1.03(0.06) & 0.04(0.06) & 0.12(0.04) &-5 & N \\
1981B$\dagger$& 1179 & -18.96(0.15) & 1.11(0.10) & 0.04(0.06) & 0.17(0.07) & 4 & HV \\
1983G  & 1497 & -18.90(0.45) & 1.30(0.10) & 0.17(0.10) & 0.23(0.09) &-2 & HV \\
1984A  & 1179 & -18.91(0.57) & 1.20(0.10) & 0.16(0.08) & 0.26(0.08) & 1 & HV \\
1989A  & 2756 & -19.17(0.25) & 1.05(0.10) & 0.10(0.10) & 0.23(0.09) & 4 & HV \\
1992A  & 1338 & -18.90(0.50) & 1.47(0.05) &-0.01(0.04) & 0.04(0.04) &-2 & HV \\
1996bo & 4893 & -18.69(0.14) & 1.28(0.06) & 0.30(0.05) & 0.40(0.05) & 5 & HV \\
1997bp & 2647 & -19.09(0.25) & 1.13(0.05) & 0.16(0.05) & 0.28(0.07) & 1 & HV \\
1997do & 3140 & -19.00(0.22) & 1.10(0.07) & 0.05(0.07) & 0.14(0.07) & 4 & HV \\
1998dh & 2766 & -19.18(0.24) & 1.22(0.04) & 0.08(0.03) & 0.20(0.04) & 5 & HV \\
1998dk & 3609 & -18.94(0.19) & 1.05(0.10) & 0.15(0.07) & 0.24(0.07) & 5 & HV \\
1998ec & 6032 & -18.61(0.13) & 1.19(0.10) & 0.28(0.08) & 0.36(0.08) & 3 & HV \\
1998ef & 5020 & -19.40(0.13) & 1.23(0.05) &-0.04(0.03) & 0.05(0.04) & 3 & HV \\
1999cc & 9452 & -18.88(0.09) & 1.48(0.10) & 0.00(0.05) & 0.06(0.06) & 5 & HV \\
1999cl & 1179 & -17.33(0.57) & 1.22(0.05) & 1.14(0.04) & 1.23(0.06) & 3 & HV \\
1999dk & 4181 & -19.13(0.16) & 1.15(0.05) & 0.08(0.05) & 0.17(0.04) & 5 & HV \\
2000fa & 6533 & -19.04(0.10) & 0.89(0.05) & 0.07(0.03) & 0.17(0.04) & 10& HV \\
2001ay & 9269 & -19.04(0.08) & 0.72(0.05) & 0.06(0.06) & 0.32(0.05) & 4 & HV \\
2001br & 6057 & -18.52(0.11) & 1.34(0.05) & 0.09(0.03) & 0.17(0.04) & 1 & HV \\
2001da & 4790 & -18.84(0.14) & 1.20(0.05) & 0.17(0.04) & 0.29(0.05) & 2 & HV \\
2001E  & 6140 & -18.41(0.11) & 1.07(0.05) & 0.50(0.05) & 0.59(0.05) & 5 & HV \\
2001en & 4478 & -18.98(0.15) & 1.28(0.05) & 0.00(0.03) & 0.06(0.05) & \nodata & HV \\
2002bf & 7418 & -19.03(0.11) & 1.00(0.10) & 0.03(0.07) & 0.17(0.06) & 3 & HV \\
2002bo & 1547 & -18.16(0.43) & 1.16(0.03) & 0.40(0.03) & 0.51(0.06) & 1 & HV \\
2002cd & 2919 & -18.18(0.22) & 1.02(0.06) & 0.63(0.04) & 0.71(0.04) & 4 & HV \\
2002cs & 4592 & -18.96(0.14) & 1.05(0.03) &-0.04(0.03) & 0.11(0.06) &-5 & HV \\
2002cu & 6889 & -18.88(0.10) & 1.42(0.05) & 0.04(0.03) & 0.12(0.04) &-5 & HV \\
2002de & 8474 & -18.82(0.08) & 1.05(0.03) & 0.09(0.03) & 0.19(0.04) & 1 & HV \\
2002dj & 2825 & -19.18(0.23) & 1.08(0.03) & 0.09(0.04) & 0.18(0.05) &-5 & HV \\
2002ef & 6846 & -18.60(0.10) & 1.08(0.05) & 0.33(0.05) & 0.38(0.04) &-2 & HV \\
2003cq & 10117& -18.79(0.08) & 1.37(0.05) & 0.07(0.07) & 0.19(0.04) & 4 & HV \\
2003hx & 1994 & -18.18(0.34) & 1.23(0.10) & 0.50(0.09) & 0.55(0.08) &-1 & HV \\
2003W  & 6334 & -19.08(0.11) & 1.08(0.05) & 0.11(0.04) & 0.22(0.07) & 5 & HV \\
2004as & 9612 & -18.85(0.07) & 1.15(0.05) & 0.10(0.04) & 0.23(0.04) &\nodata& HV \\
2004bk & 7225 & -19.24(0.10) & 1.18(0.08) & 0.03(0.06) & 0.09(0.05) & 3 & HV \\
2004dt & 5644 & -19.25(0.12) & 1.12(0.05) & 0.00(0.04) & 0.11(0.05) & 1 & HV \\
2004ef & 8931 & -18.76(0.08) & 1.42(0.05) & 0.05(0.04) & 0.09(0.04) & 3 & HV \\
2005A  & 5502 & -17.39(0.12) & 1.28(0.03) & 1.01(0.01) & 1.07(0.04) & 4.5& HV \\
2006ac & 7180 & -18.97(0.10) & 1.23(0.08) & 0.12(0.07) & 0.19(0.05) & 3 & HV \\
2006bq & 6432 & -18.71(0.11) & 1.57(0.06) & 0.16(0.04) & 0.17(0.04) &-2 & HV \\
2006br & 7657 & -17.15(0.10) & 1.55(0.15) & 1.03(0.07) & 1.05(0.06) & 3 & HV \\
2006bt & 9737 & -18.93(0.07) & 1.12(0.05) & 0.14(0.05) & 0.23(0.04) & 0 & HV \\
2006cp & 6991 & -19.18(0.10) & 1.13(0.05) & 0.10(0.04) & 0.20(0.04) & 5 & HV \\
2006ef & 5102 & -18.90(0.13) & 1.37(0.05) & 0.01(0.04) & 0.06(0.04) &-1 & HV \\
2006ej & 5801 & -18.85(0.12) & 1.40(0.12) &-0.01(0.06) & 0.03(0.04) & 5 & HV \\
2006gr & 10037& -18.92(0.08) & 0.93(0.05) & 0.09(0.06) & 0.19(0.05) & 3 & HV \\
2006le & 5178 & -19.48(0.13) & 0.88(0.05) &-0.10(0.04) & 0.02(0.04) & 3 & HV \\
2006os & 9624 & -18.45(0.07) & 1.44(0.07) & 0.30(0.04) & 0.41(0.05) & \nodata & HV \\
2006sr & 6960 & -18.88(0.11) & 1.28(0.09) & 0.07(0.06) & 0.09(0.05) & 6 & HV \\
2006X$\dagger$  & 957  & -16.98(0.14) & 1.31(0.05) & 1.35(0.04)  & 1.42(0.05) & 4 & HV \\
2007bd & 9581 & -19.11(0.07) & 1.35(0.12) & 0.00(0.05) & 0.05(0.04) & 1 & HV \\
2007co & 7963 & -18.91(0.09) & 1.13(0.05) & 0.08(0.04) & 0.16(0.05) & \nodata & HV \\
2007gi & 1824 & -19.06(0.36) & 1.37(0.05) & 0.15(0.05) & 0.16(0.05) & -3& HV \\
2007le & 2041 & -18.78(0.32) & 1.02(0.05) & 0.37(0.04) & 0.45(0.04) & 5 & HV \\
2007qe & 7200 & -19.09(0.10) & 1.02(0.05) & 0.10(0.05) & 0.22(0.04) & \nodata & HV \\
2007sr & 1757 & -19.31(0.38) & 1.16(0.07) & 0.11(0.08) & 0.23(0.07) & 9 & HV \\

\enddata
\tablenotetext{$\ast$} {Uncertainties are $1\sigma$. See text for a
discussion of recession velocities. Host-galaxy ``T'' types are from
RC3.} \tablenotetext{$\dagger$}{SN~Ia with Cepheid distance.}
\tablenotetext{$\ddagger$}{The $\Delta m_{15}$ value has been
corrected for the reddening effect (Phillips et~al. 1999).}

\end{deluxetable}



\begin{thebibliography}{}
\bibitem [Altavilla et~al. (2007)]{alta07} Altavilla, G., et~al. 2007, \aap, 475, 585
\bibitem [Barbon et~al. (1990)]{bar90} Barbon, R., et~al. 1990,\aap, 237, 79
\bibitem [Benetti et~al. (2004)]{ben04} Benetti, S., et~al. 2004, \mnras, 348, 261
\bibitem [Benetti et~al. (2005)]{ben05} Benetti, S., et~al. 2005, \apj, 623, 1011 (B05)
\bibitem [Blondin et~al. (2009)]{blon09} Blondin, S., et~al. 2009, \apj, 693, 207
\bibitem [Branch et al. (2009)]{bran09} Branch, D., Dang, L., \& Baron, E.
  2009, \pasp, 121, 238
\bibitem [Branch et~al. (1983)]{bran83} Branch, D., et~al. 1983, \apj, 270, 123
\bibitem [Branch et~al. (2003)]{bran03} Branch, D., et~al. 2003, \aj, 126, 1489
\bibitem [Branch et~al. (2006)]{bran06} Branch, D., et~al. 2006, \pasp, 118, 560
\bibitem [de Vaucouleurs et al. (1991)]{devauc91} de Vaucouleurs, G., et al., 1991, Third Reference
  Catalogue of Bright Galaxies (New York: Springer-Verlag) (RC3)
\bibitem [Dominguez et~al. (2001)]{dom01} Dominguez, I., H\"{o}flich, P., \& Starniero,
  O. 2001, \apj, 557, 279
\bibitem [Elias-Rosa et~al. (2006)]{elis06} Elias-Rosa, N., et~al. 2006, \mnras, 369, 1880
\bibitem [Filippenko (1992a)]{fili92a} Filippenko, A. V., et~al. 1992a, \apj, 384, L15
\bibitem [Filippenko (1992b)]{fili92b} Filippenko, A. V., et~al. 1992b, \aj, 104, 1543
\bibitem [Filippenko (1997)]{fili97}Filippenko, A. V. 1997, \araa, 35, 309
\bibitem [Filippenko et~al. (2001)]{fili01} Filippenko, A. V., et al. 2001,
in Small Telescope Astronomy on  Global Scales, ed. B.
Paczy\'{n}ski, W.-P. Chen, \& C. Lemme (San Francisco: ASP), 121
\bibitem [Garavini et~al. (2007)]{gara07} Garavini, G., et~al. 2007, \aap, 471, 527
\bibitem [Gerardy et~al. (2004)]{ger04} Gerardy, C., et~al.  2004, \apj, 607, 391
\bibitem [Goobar et~al. (2008)]{goob08} Goobar, A. 2008, \apj, 686, L103
\bibitem [Hachinger et~al. (2006)]{hac06} Hachinger, S., et~al. 2006, \mnras, 2006, 370, 299
\bibitem [Hamuy et~al. (2002)]{ham02} Hamuy, M., et~al. 2002, \aj, 124, 417
\bibitem [Hicken et~al. (2009)]{hick09} Hicken, M., et~al. 2009 (arXiv:0901.4787)
\bibitem [Hillebrandt \& Niemeyer (2000)]{hn00} Hillebrandt, W., \& Niemeyer, J. C.
  2000, \araa, 38, 191
\bibitem [Jha et al. (2007)]{jha07} Jha, S., Riess, A. G., \& Kirshner, R. P.  2007,
  \apj, 659, 122
\bibitem [Komatsu et al. (2009)]{Koma09} Komatsu., E., et al. 2009, \apjs, 180, 330
\bibitem [Kotak et al. (2005)]{kot05} Kotak, R., et al. 2005, \aap, 436, 1021
\bibitem [Li, et~al. (2003)]{li03} Li, W., et~al. 2003, \pasp, 115, 453
\bibitem [Matheson et~al. (2008)]{math08} Matheson, T., et~al. 2008, \aj, 135, 1598 (M08)
\bibitem [Nugent et al. 1995]{nug95} Nugent, P., et al. 1995, \apj, 455, L147
\bibitem [Pastorello et~al.(2007)]{past07} Pastorello, A., et~al. 2007, \mnras, 377, 1531
\bibitem [Patat et~al. (1996)]{pat96} Patat, F., et~al. 1996, \mnras, 278, 111
\bibitem [Patat et~al. (2007)]{pat07} Patat, F., et~al. 2007, Science, 317, 924
\bibitem [Perlmutter et~al. (1999)]{perl99} Perlmutter, S., et~al. 1999, \apj, 517, 565
\bibitem [Phillips (1993)]{phi93} Phillips, M. M. 1993, \apj, 413, L105
\bibitem [Phillips (1999)]{phi99} Phillips, M. M., et al. 1999, \aj, 118, 1766
\bibitem [Pignata et~al.(2008)]{pign08} Pignata, G., et~al. 2008, \mnras, 388, 971
\bibitem [Riess et al. (1998)]{rie98} Riess, A. G., et al. 1998, \aj, 116, 1009
\bibitem [Schlegel et al. (1998)]{sfd98} Schlegel, D. J., Finkbeiner, D. P., \&
Davis, M. 1998, \apj, 500, 525
\bibitem [Stanishev et~al. (2007)]{stan07} Stanishev, V., et~al. 2007, \aap, 469, 645
\bibitem [Suntzeff 1996]{suntz06} Suntzeff, N. B. 1996, in Supernovae and
Supernova Remnants, ed. R. McCray \& Z. Wang (Cambridge: Cambridge Univ. Press), 41
\bibitem [Tanaka et~al. (2008)]{tan08} Tanaka, M., et~al. 2008, \apj, 677, 448
\bibitem [Timmes et al. (2003)]{Timmes03} Timmes, F. X., Brown, E. F., \& Truran,
   J. W. 2003, \apj, 590, L83
\bibitem [L. Wang et~al. (2005)]{wang_l05} Wang, L., et al. 2005, \apj, 635, L33
\bibitem [L. Wang et~al. (2006)]{wang_l06} Wang, L., et~al. 2006, \apj, 653, 490
\bibitem [X. Wang et~al. (2005)]{wang_x05} Wang, X., et~al. 2005, \apj, 620, L87
\bibitem [X. Wang et~al. (2006)]{wang_x06} Wang, X., et~al. 2006, \apj, 645, 488
\bibitem [Wang et~al. (2008a)]{wang08a} Wang, X., et~al. 2008a, \apj, 675, 626 (W08a)
\bibitem [Wang et~al. (2008b)]{wang08b} Wang, X., et~al. 2008b, \apj, 677, 1060
\bibitem [Wang et~al. (2009)]{wang09} Wang, X., et~al. 2009, \apj, 697, 380
\bibitem [Yamanaka et al. (2009)]{yam09} Yamanaka, M., et al. 2009, \pasj, in press (arXiv:0904.2763)
\end{thebibliography}
\end{document}